\documentclass[aps,prl,twocolumn,showpacs,superscriptaddress]{revtex4-1}  % for review and submission

\usepackage{upgreek}
\usepackage{graphicx,import}  % needed for figures
\usepackage{dcolumn}   % needed for some tables
\usepackage{bm}        % for math
\usepackage{color}
\usepackage{amssymb}
\usepackage{amsmath}
\usepackage{braket}
\usepackage{siunitx}
\usepackage{array}

\begin{document}
\widetext
% the following line is for submission, including submission to the arXiv!!
%\hspace{5.2in} \mbox{Fermilab-Pub-04/xxx-E}

\title{Precision measurement of the ionization energy of a single trapped $^{40}$Ca$^+$ ion by Rydberg series excitation}
%\input author_list.tex       % D0 authors (remove the first 3 lines
                             % of this file prior to submission, they
                             % contain a time stamp for the authorlist)
                             % (includes institutions and visitors)

\author{J. Andrijauskas} \affiliation{QUANTUM, Johannes Gutenberg-Universit\"at Mainz, Staudinger Weg 7, 55128 Mainz, Germany}
\author{J. Vogel} \affiliation{QUANTUM, Johannes Gutenberg-Universit\"at Mainz, Staudinger Weg 7, 55128 Mainz, Germany}
\author{A. Mokhberi} \affiliation{QUANTUM, Johannes Gutenberg-Universit\"at Mainz, Staudinger Weg 7, 55128 Mainz, Germany}
\email{arezoo.mokhberi@uni-mainz.de}
\author{F. Schmidt-Kaler} \affiliation{QUANTUM, Johannes Gutenberg-Universit\"at Mainz, Staudinger Weg 7, 55128 Mainz, Germany}
\affiliation{Helmholtz-Institut Mainz, 55128 Mainz, Germany}
\date{\today}

\begin{abstract}
A complete set of spectroscopic data is indispensable when using Rydberg states of trapped ions for quantum information processing. We carried out Rydberg series spectroscopy for $nS_{1/2}$ states with $38 \leq n \leq 65$ and for $nD_{5/2}$ states with $37\leq n \leq 50$ on a single trapped $^{40}$Ca$^+$ ion. From a nonlinear regression to resonance frequencies, we determined the ionization energy of 2 870 575.582(15)~GHz, measured 60 times more accurately as compared to the accepted value and contradicting it by 7.5 standard deviations. We confirm quantum defect values of $\delta_{S_{1/2}}=1.802995(5)$ and $\delta_{D_{5/2}}=0.626888(9)$ for $nS_{1/2}$ and $nD_{5/2}$ states respectively, which allow for unambiguous addressing of Rydberg levels of Ca$^+$ ions.
\end{abstract}

%\pacs{37.10.Ty}% Ion trappings.

\maketitle

Single trapped ions and ion crystals are employed for high-precision spectroscopy and frequency standards~\cite{ludlow15a}, for fundamental quantum optics experiments, for quantum simulation~\cite{blatt08a} and quantum computing~\cite{bermudez17a}. This success is based on the near-to-perfect control of their electronic and vibrational quantum states. If the valence electron of ions is excited to the states with high principal quantum numbers $n$, called Rydberg states, ions feature additional and exotic properties~\cite{mokhberi20a}. Similar to the neutral Rydberg atom counterpart, where experiments successfully demonstrate quantum entanglement~\cite{isenhower10a, wilk10a} and quantum simulation~\cite{gross17a,browaeys2020}, Rydberg ions can be exploited for their tunable, strong dipolar interactions and large polarizabilities~\cite{feldker15a, higgins17a, higgins19a}. The combination of advantages of trapped ions with those arising from exotic Rydberg properties are opening novel pathways to generate multi-particle entanglement~\cite{vogel19a, zhang20a}, quantum simulation and computation~\cite{mueller08a, li13a}.

However, to fully establish the novel quantum platform of Rydberg ions, a reliable and complete set of accurate spectroscopic data is essential. Our endeavor for precision spectroscopy of Rydberg levels uses a single cold trapped ion, thus minimizing systematic line shifting effects.
We discuss the results for eighteen $nS_{1/2}$ and fourteen $nD_{5/2}$ Rydberg transitions in $^{40}$Ca$^+$ to determine the ionization energy of $^{40}$Ca$^+$ with high accuracy and extending high resolution spectroscopy in Rydberg neutral gases \cite{peper19a} to Rydberg ions. We first describe the experimental setup, and discuss our spectroscopy sequence. Spectra are analyzed, line-shape models are discussed, to determine an error budget for the value of ionization energy and quantum defect.

\begin{figure}[h!]
\resizebox{0.48\textwidth}{!}{
\includegraphics{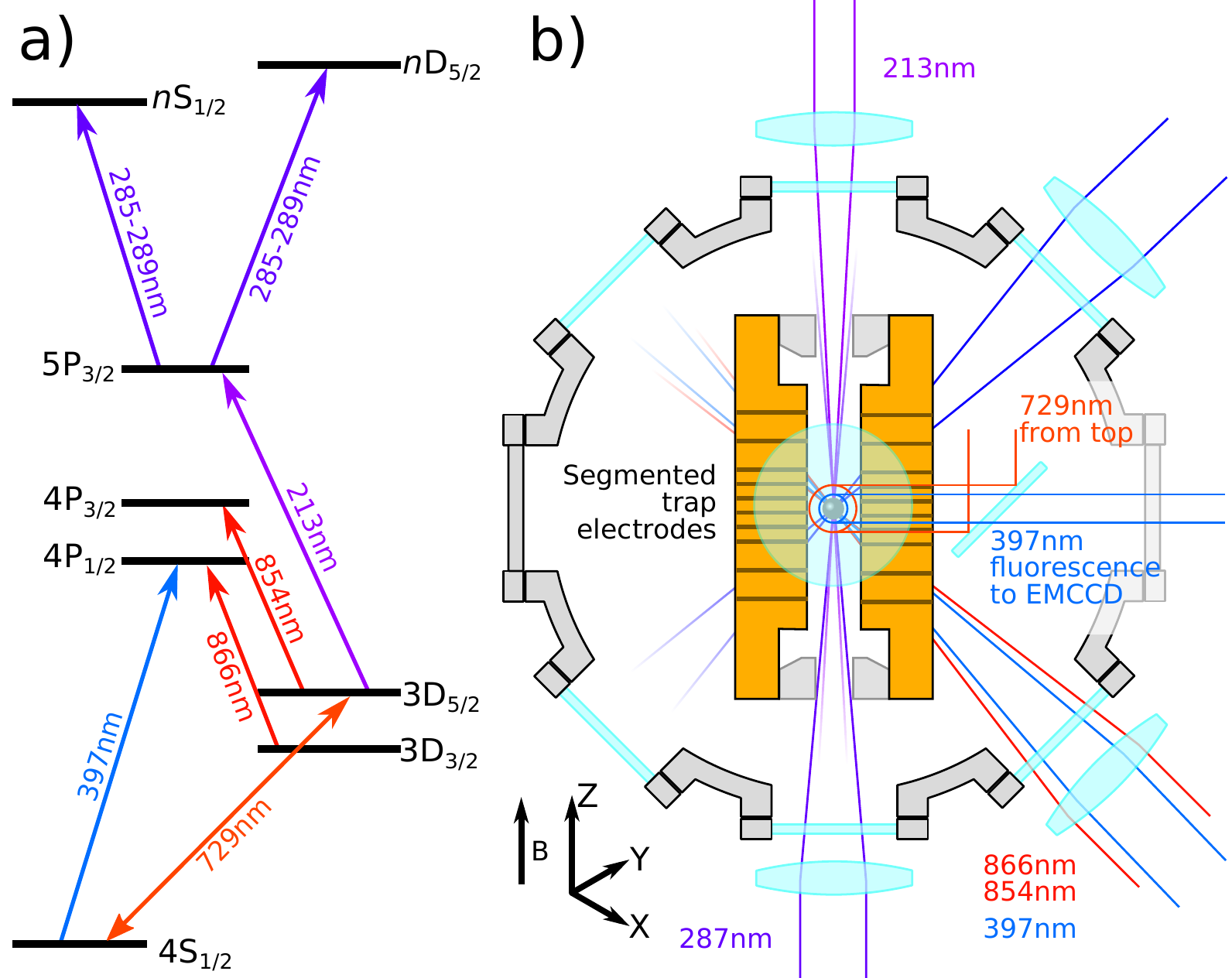}}
\caption{\label{fig:chamberandenergy} a) Diagram of relevant energy levels in $^{40}$Ca$^{+}$ with transition wavelengths for laser cooling, Rydberg excitation and state detection, Zeeman substructure omitted. b) Sketch of setup with two out of four segmented trap electrodes (yellow) visible in top view. UV-beams near 213~nm and 287~nm enter counter-propagating through holes in the endcap electrodes (grey). Viewports and lenses (light blue) allow for optical access of beams at 397~nm, 866~nm, 854~nm. From the top, a beam near 729~nm is focused by an objective lens with 0.27 N.A. on the ion. Fluorescence light scattered by the ion near 397~nm is collected also by this objective, steered by dichroic mirror on an EMCCD. A homogeneous magnetic field of 5.573(3)$\times 10^{-4}$~T is applied in the z-direction.}
\end{figure}

\begin{figure*}
\includegraphics[width=\textwidth]{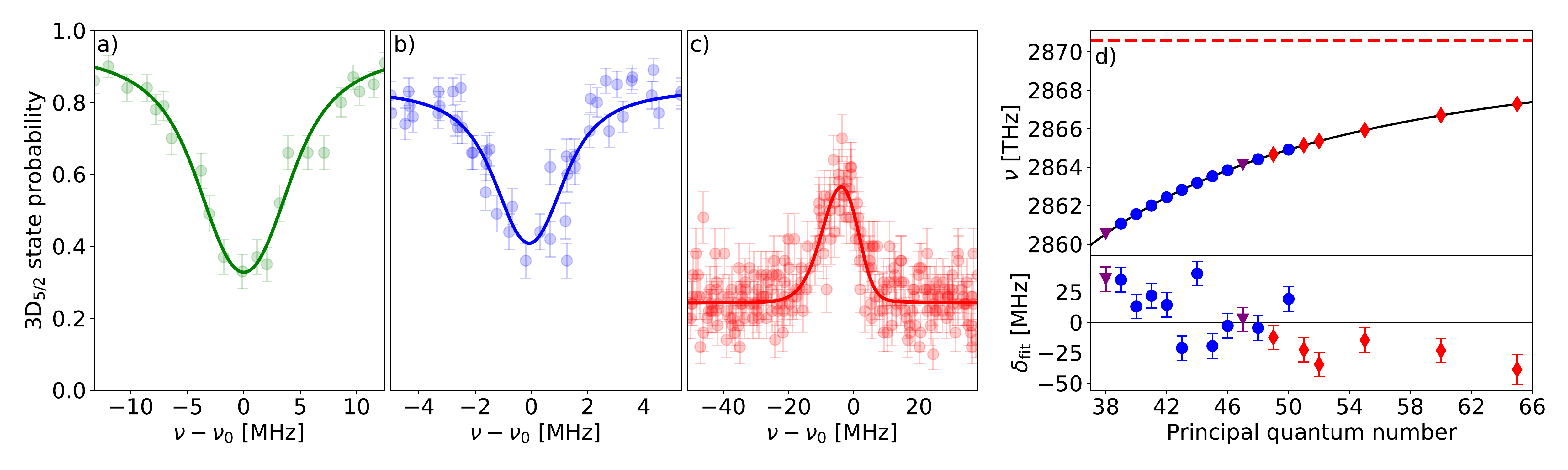}
\caption{(a) $3D_{5/2}$($m_J=-5/2$)$\rightarrow5P~{_{3/2}}$($m_J=-3/2$) excitation spectrum, (b) $3D_{5/2}~$($m_J=-5/2$)$\rightarrow38S_{1/2}$($m_J=-1/2$) excitation spectrum and (c) $3D_{5/2}$($m_J=-5/2$)$\rightarrow 55S_{1/2}$($m_J=-1/2$) quantum interference (QI) spectrum. Measurements~(dots) and fits~(solid lines) are shown. In (a) and (c), error bars depict quantum projection noise for 50 sequence repetitions and in (b) for 100 ones, see details in text. (d) Transition frequencies of the S states fitted to Eqn.~\ref{eq:ritz} and its fit residuals. The data were obtained from two-photon excitation spectra (blue circles), QI spectra (red diamonds), and from both (purple triangles), error bars show the statistical uncertainty.}
\label{fig:lines_and_fit}
\end{figure*}

%\textit{Experimental setup.--}
We employ a segmented linear Paul trap driven with a radiofrequency~(RF) voltage to generate the confining potential in $x-y$ plane and static voltages applied to endcap and segmented electrodes for the z-confinement of $^{40}$Ca$^+$~(Fig.~\ref{fig:chamberandenergy}(b)). Under typical operation conditions with a RF drive of $\Omega_{RF} = 2\pi \times \SI{14.56}{\mega \hertz}$ and amplitude of $V_{pp} = \SI{580}{\volt}$ we obtain secular trapping frequencies of $\omega_{x,y,z}=2\pi \times \{ 950,1270,555\}$~kHz. A single ion is loaded into the trap by photoionizing from a neutral Calcium beam~\cite{gulde01a}. To avoid contamination of the trap electrodes, which may cause parasitic electric fields and surface charges, we use a separate loading zone and transport the ion then to the shielded central zone. For Doppler cooling and optical pumping of the ion, we employe lasers at \SIlist{397;866;854;729}{\nano \meter}, driving the $4S_{1/2}\rightarrow 4P_{1/2}$, the $3D_{3/2}\rightarrow 4P_{1/2}$, the $3D_{5/2}\rightarrow 4P_{3/2}$ and the $4S_{1/2}\rightarrow 3D_{5/2}$ transitions, respectively (Fig.~\ref{fig:chamberandenergy}(a)). Laser-induced fluorescence near 397~nm was collected using an objective system and imaged into a electron-multiplying charge-coupled device (EMCCD) camera.

We excite Rydberg states in a two-step process with ultra-violet~(UV) light near 213~nm and 287~nm wavelength, addressing the $3D_{5/2}\rightarrow 5P_{3/2}$ and $5P_{3/2}\rightarrow nS_{1/2}$ or $5P_{3/2}\rightarrow nD_{5/2}$ transitions, respectively~(Fig.~\ref{fig:chamberandenergy}(a)). The UV beams stem from frequency quadrupled lasers at \SIlist{852;1144}{\nano \meter}, both stabilized to reference cavity for a linewidth of 1.0(0.5)~MHz. UV beams are counter-propagating, pass through the holes in endcap electrodes and impinge the ion. To align them we detect their induced AC Stark shift on a $4S_{1/2} - 3D_{5/2}$ superposition state. Beam waists are about 7.5(1.5)~$\mu$m and 25(5)~$\mu$m and up to 12~mW (only a few $\mu$W used for spectroscopy) and 80~mW for the 213~nm and the 287~nm beam, respectively. Wavelengths of the \SIlist{852;1144}{\nano \meter} sources are measured by a High Finesse WS8-2 wavemeter, calibrated to the narrow $4S_{1/2} \rightarrow 3D_{5/2}$ clock transition of $^{40}$Ca$^+$~\cite{chwalla09a}.

We carried out the spectroscopy sequence: (i) Doppler cooling using 397~nm, 866~nm and 854~nm laser beams for 12~ms, and  detecting ion fluoresence to asure the properly cooled single ion, followed by (ii) frequency-selective optical pumping in the $4S_{1/2}$($m_J=-1/2$) with 99\% purity \cite{roos06a} by applying 8~pulses of 20~$\mu$s length near 729~nm, which drive the $4S_{1/2}$($m_J=+1/2$)$ \rightarrow 4D_{5/2}$($m_J=-3/2$) transition, each follwed by a quench using radiation near 854~nm for 5$\mu$s. (iii) Rapid-adiabatic passage to the $3D_{5/2}$($m_J=-5/2$) state using a 200~$\mu$s pulse length and 200~kHz frequency sweep at 729~nm. This initializes with 95\% efficiency in the $3D_{5/2}$ state with 1.168~s lifetime~\cite{kreuter05a}. (iv) Exciting Rydberg states via the intermediate $5P_{3/2}$($m_J = -3/2$) level using two beams near 213~nm and 287~nm. Both beams are circularly polarized ($\sigma^+_{213} / \sigma^+_{287}$ for $S$ states, and $\sigma^+_{213} / \sigma^-_{287}$ for $D$ states) to drive the electric dipole-allowed transitions with $\Delta m_J=\pm1$. Excited Rydberg states exhibit lifetimes in the order of a few $\mu$s and decay predominantly into the $S_{1/2}$ ground state. Therefore, successful Rydberg excitation was detected from a population reduction in the $3D_{5/2}$ state. (v) For this we switch on laser radiation at 397~nm and 866~nm for 3~ms and observe laser-induced fluorescence on the EMCCD. The detection of UV-fluorescence indicates a successful Rydberg excitation, while we observe no fluorescence if the ion remained in the $3D_{5/2}$. The experimental sequence from (i) to (v) is repeated, typically in 50 cycles. The results in (v) are averaged, where we exclude data without successful initial ion detection in (i). Finally, the scanning parameter, e.g. the laser frequency is stepped to the next value.

% measure
In preparation for the Rydberg excitation, we investigated the $3D_{5/2}$ to $5P_{3/2}$ transition. Only the laser near 213~nm was used for 1~ms in step (iv), depleting the $3D_{5/2}$ population if resonant with the transition (Fig.~\ref{fig:lines_and_fit}(a)). Now, to measure Rydberg spectra we employ two different methods: Either we scan the detuning $\Delta_{287}$ of the second laser over the resonance, while we chose a fixed detuning of $\Delta_{213}\sim$100~MHz, thus mitigating decay from the short-lived intermediate level $5P_{3/2}$. Both laser fields are switched on at step (iv) of the sequence for a 1~ms duration. Successful Rydberg excitation depletes the population in $D_{5/2}$, and the resonance appears as a dip~(Fig.~\ref{fig:lines_and_fit}(b)). Alternatively, we fix $\Delta_{213}$ resonant for the $5P_{3/2}$ transition and vary the detuning $\Delta_{287}$. Because of scattering events to the short-lived  $5P_{3/2}$ state ($\tau \sim 34.8(7)$~ns~\cite{safronova11a}) the population in $D_{5/2}$ is transferred to the ground $4S_{1/2}$ state. However, as the second laser approaches a Rydberg resonance, the levels couple and the $5P_{3/2}$ becomes energetically shifted. Thus, the 213~nm light field is off-resonant with such intermediate dressed state and population remains in the $3D_{5/2}$ state, and hence the resonance appears as peak~(Fig.~\ref{fig:lines_and_fit}(c)). This excitation method is understood from quantum interference (QI), similar as in electromagnetically induced transparency for an atomic three-level system~\cite{mohapatra07a, eschner03a}.

%ion loss FSK
Ions in highly excited Rydberg states may be unstable from ionization due to the electric field of the trap~\cite{mueller08a} or from transitions driven by black body radiation~\cite{beterov09a}. We observe that such doubly ionized Ca$^{2+}$ remains trapped, inaccessible by the optical fields~\cite{bachor16a}. We recorded such events and normalized the rate by the rate of successful excitations to the Rydberg states. Losses are increasing from 1.8(4)~10$^{-3}$ for $n=38$ to 8.0(2.8)~10$^{-3}$ for $n=52$, see table~\ref{tab:data}, the  rate follows a power law scaling with $n^{\num{5.7(9)}}$, see supplementary material A. From the match with the expected power law $n^{\num{5}}$ in Ref.~\cite{beterov09a} we conjecture black-body radiation as the dominant loss effect. For QI spectroscopy no ion loss is observed as it involves Rydberg-dressing only. Background gas collisions losses occur at time scales of hours and are negligible.

\begin{table*}
  \caption{Rydberg transition frequencies including uncertainties from systematic and statistical errors, full width at half maximum linewidths (FWHM) and ion loss probabilities for $nS_{1/2}$ and $nD_{5/2}$ states for a single trapped $^{40}$Ca$^+$ ion. Linewidths extracted from QI spectra are marked with asterisks.}
  \label{tab:data}
  \begin{center}
  {\renewcommand{\arraystretch}{1.4}
  \begin{tabular}{m{0.04\textwidth} m{0.18\textwidth} m{0.16\textwidth} m{0.08\textwidth} m{0.18\textwidth} m{0.12\textwidth}}
  \hline \hline
  $n$ & $\nu_{n{\rm S}_{1/2}}$~[\si{\giga \hertz}] & FWHM~[\si{\mega \hertz}] & $P_{\text{loss}} [\%]$ & $\nu_{n{\rm D}_{5/2}}$~[\si{\giga \hertz}] & FWHM~[\si{\mega \hertz}] \\
  \hline
  37  & --                  & --                  & --       & 2 860 629.105(83)  & 3.1(0.6)/5.4(0.4)*   \\
  38  & 2 860 531.993(83)   & 3.3(0.7)/22.9(0.7)* & 0.18(4)  & 2 861 154.272(83)  & 9.3(0.7)*          \\
  39  & 2 861 064.764(83)   & 3.3(0.4)            & 0.22(4)  & 2 861 638.895(83)  & 4.8(0.3)/9.5(1.7)* \\
  40  & 2 861 556.219(83)   & 3.9(0.8)            & 0.31(7)  & 2 862 087.091(83)  & 3.7(0.7)/6.8(0.9)*   \\
  41  & 2 862 010.570(83)   & 3.0(0.6)            & 0.46(11) & 2 862 502.389(83)  & 9.5(0.9)*          \\
  42  & 2 862 431.419(83)   & 5.2(0.7)            & 0.38(9)  & 2 862 887.930(83)  & 15(3)*           \\
  43  & 2 862 821.966(83)   & 3.7(0.8)            & 0.60(17) & 2 863 246.513(83)  & 4.3(0.7)*          \\
  44  & 2 863 185.171(83)   & 3.9(0.9)            & 0.41(13) & 2 863 580.555(83)  & 6.2(0.7)*        \\
  45  & 2 863 523.328(83)   & 6.0(0.8)            & 0.43(11) & 2 863 892.284(83)  & 5.8(1.3)*        \\
  46  & 2 863 838.864(83)   & 3.5(0.6)            & 0.79(21) & 2 864 183.640(83)  & 6.8(0.8)*          \\
  47  & 2 864 133.678(83)   & 6.1(1.2)/13.1(0.5)* & 0.69(21) & 2 864 456.336(83)  & 6.5(1.4)*        \\
  48  & 2 864 409.543(83)   & 5.2(0.8)            & 0.81(28) & 2 864 711.949(83)  & 5.8(0.7)*          \\
  49  & 2 864 668.058(83)   & 9.7(0.6)*           & --       & 2 864 951.927(83)  & 11(5)*           \\
  50  & 2 864 910.688(83)   & 6.5(0.7)            & 0.80(28) & 2 865 177.383(83)  & 7.5(1.0)*        \\
  51  & 2 865 138.602(83)   & 13.1(0.6)*          & --       & --                 & --               \\
  52  & 2 865 353.058(83)   & 18.8(0.6)*          & --       & --                 & --               \\
  55  & 2 865 925.507(83)   & 12.2(0.6)*          & --       & --                 & --               \\
  60  & 2 866 690.197(83)   & 16.2(0.6)*          & --       & --                 & --               \\
  65  & 2 867 280.663(85)   & 16.3(0.9)*          & --       & --                 & --               \\
  \hline \hline
  \end{tabular}
  }
  \end{center}
\end{table*}

%analyses
We fitted the $3D_{5/2}\rightarrow 5P_{3/2}$ transition resonance to a Voigt profile (Fig.~\ref{fig:lines_and_fit}(a)). The theoretical profile is taking into account the theory value of the natural linewidth~\cite{safronova11a} of $2 \pi \times 4.57(9)$~MHz, and an additional laser linewidth broadening of 1.0(0.5)~MHz. The Doppler broadening extracted from the fit is 4.5(1.1)~MHz. The Zeeman shift for the $3D_{5/2}$($m_J=-5/2$)$\rightarrow 5P_{3/2}$($m_J = -3/2$) was measured to be 7.80(1)~MHz using resolved sideband spectroscopy in the 4S$_{1/2}\rightarrow$3D$_{5/2}$ transition. We determine a centroid frequency of the $5P_{3/2}$ state at 1~817~081.257(30)~GHz. The error budget includes a 13~MHz systematic uncertainty due to the wavemeter, a 10~MHz calibration frequency uncertainty and 7~MHz statistical uncertainty from the fitting, see Supplementary Material. Our measurement is 10 times more accurate and with 2.7~$\sigma$ (+800~MHz) above the accepted value of 1~817~080.5(0.3)~GHz~\cite{sugar79a}.
%of 60,611.28~cm$^{-1}$~\cite{sugar79a}. CHECK NUMBERS FOR CENTROID}.

% Ich möchte dieses weglassen:  "and the noise added by the fiber coupler for the calibration laser light, " ausserdem, das wäre eher in der 10MHz calibration beinhaltet ?? $\Delta \omega=2 \pi \times 4.6$(0.1)~MHz

% Doppler
% Micromotion
% S / D states quadrupolar shift
% AC Stark
% Equ
% Phonon number
To analyse the obtained Rydberg spectra, we complete the line model by taking into account the polarizability of high $n$-states and quadrupole couplings to the dynamic Paul trap potential~\cite{mokhberi20a}. In these calculations, the following systematic shifts and broadening mechanisms were considered: The Doppler broadening effect is largely suppressed as both exciting UV beams are counter-propagating. We calculate the residual Doppler broadening to be about 2.1(0.1)~MHz for a wavepacket in thermal motion with $T\approx 2.6$~mK, while the relativistic Doppler shift is fully negligible. The natural linewidths of the excited Rydberg states~\cite{glukhov13a} are about an order of magnitude smaller as compared to the combined laser linewidth of both UV beams which account for 2(1)~MHz. Before each run of data acquisition, we minimize the ion's excess micromotion, occuring when the ion is displaced from the trap center~\cite{berkeland98a}. This procedure requires a precise measurement and compensation of parasitic static electric fields. For this, we used resolved sideband spectroscopy on the 4$S_{1/2}\rightarrow$ 3D$_{5/2}$ transition~\cite{mokhberi19a} and determined E$_{rms} \leq 40$(10)~V/m, with the error estimated from the ion positioning in the $xy$-plane. Owing to the polarizibility of Rydberg states~\cite{feldker15a, higgins19a}, such electric field may lead to a quadratic Stark shift. In case of the $65S$, the state with the highest electric susceptibility interrogated in this work, this shift accounts for $-13$(6)~MHz. A further important effect is due to the AC Stark shift from coupling the Rydberg electron to the RF electric field. This effect is present even when parasitic electric fields are perfectly compensated and leads to a change of the ion motional frequency $\omega^{\prime}_{x,y} \approx \sqrt{\omega^2_{x,y} - 2\mathcal{\alpha}\gamma^2/M }$, with secular trapping frequencies $\omega_{x,y}$ for the electronic ground-state, the ion mass $M$, the excited state polariziability $\alpha$ and the gradient $\gamma$ of the RF field. Using perturbation theory, the energy shift is proportional to the polarizability of the excited states~\cite{mueller08a}. The effect manifests itself as a phonon-dependent shift and causes asymmetric lineshapes~\cite{higgins17a}. Note that this energy shift is noticeable for the motional modes in the $xy$-plane because of the comparably strong RF confinement along the radial directions, while it is about two orders of magnitude reduced for axial modes along $z$-direction, under typical operation conditions. For each set of radial Fock states with phonon numbers $n_x$, $n_y$, the resonance shifts by $\delta \omega_{sec} = n_x \delta \omega_x + n_y \delta \omega_y$, where $\Delta \omega$ is $\omega^{\prime}_{x,y}-\omega_{x,y}$. We sum up Voigt profiles with these shifts over the respective probabilities to find $n$ in a thermal phonon distribution $p_n=n_{th}^n exp(-n_{th})/n!$ with average number $n_{th}$  and used calculated values for $\alpha$~\cite{kamenski14a}. While we vary ion temperature, we keep the common linewidth of both excitation lasers at 2(1)~MHz and fit the corresponding line profile to the data, resulting in average ion temperatures of 2.6(0.2)~mK. As polarizability scales with the principal quantum number as $n^7$, the line asymmetry and broadening are more pronounced for higher Rydberg states, see Table~\ref{tab:data}.

Having corrected for all the above shifts and taken into account the frequency shift $+15.60$(1)~MHz due to the Zeeman splitting, we determined the center frequencies and linewidths as presented in Table~\ref{tab:data}. A similar procedure was applied in analyses of $n$D lines. The absolute accuracy for measuring the Rydberg resonances is limited by the WS8-2 wavemeter and its frequency calibration, which results in a total systematic error of \SI{73}{\mega \hertz} for laser frequencies near 213~nm and 287~nm wavelength, see supplementary material B. The statistical uncertainties due to the wavemeter calibration and the line fitting account for 10~MHz. For Rydberg spectra obtained using the QI method, we modelled the resonance using an additional Lorenzian broadening that scales with the coupling strength between the $5P_{3/2}$ and Rydberg states~\cite{fleischhauer05a}. The Rydberg frequencies are shown in Table~\ref{tab:data}.

%systematic shift of the Rydberg resonances, which should be distinguished from the above effect, is the
%Note that the static field at the ion equilibrium position and thus DC Stark shift are always zero.
%Other sources of systematic shifts were estimated to be neglectable in comparison with the accuracy of measurements.
%been significantly suppressed in our measuremnets as compared to the previous work~\cite{feldker14a}
%Figure~\ref{fig:lines_and_fit}(b,c) shows spectra measured using TFE and QI methods.

% accuracy discussion of shifts
% wavemeter accuracy
% total -> Tab. 1 / Fig 3

%Determination of the ionization energy and discussions
To determine the ionization energy $I^{++}$ and quantum defect parameters, we fitted the transition energies to the Rydberg-Ritz formula~\cite{ritz1908a}:
\begin{equation}
\begin{aligned}
E_{n,l,j}=& I^{++}-\frac{Z^{2} R^{*}}{(n-\mu(E))^{2}} +
\\ &\frac{Z^{4} \alpha ^{2}R^{*}}{(n-\mu(E))^{3}}{\Big [}\frac{3}{4(n-\mu(E))}-\frac{1}{(j+1/2)}{\Big ]}
\label{eq:ritz}
\end{aligned}
\end{equation}
Here, $Z$ (=$+2$) is the charge of the ionic core and $\mu(E)$ denotes the quantum defect. Using recommended fundamental constants~\cite{nistasd18} and the Ca$^{+}$ mass~\cite{IUPACmass}, we calculated the reduced Rydberg constant $R^{*}=$3~289~796.799~59(2)~GHz.
The energy dependence of the quantum defect $\mu(E)$ can be approximated by a truncated Taylor expansion:
\begin{equation}
\begin{aligned}
\mu(E_{n,l,j})\approx &\mu ^{0}_{l,j}(I^{++})+\frac {\partial \mu}{\partial E_{n,l,j}}(E_{n,l,j}-I^{++})
\\ = &\mu ^{0}_{l,j}(I^{++})-\frac {\partial \mu}{\partial E_{n,l,j}} \frac{R^{*}}{(n-\mu^{1}_{l,j})^{2}},
\label{eq:qd}
\end{aligned}
\end{equation}
with coefficients $\mu ^{0}_{l,j}$, $\mu ^{1}_{l,j}$, and ${\partial \mu}/{\partial E_{n,l,j}}$.
For the states with $n \geq 37$, the contribution of higher order terms extending Eqn.~\ref{eq:qd} is smaller than the measurement uncertainties, while the second term contribution is in the order of experimental uncertainties. Thus, energy-dependent coefficients have to be extracted using also $nS_{1/2}$ states with $5 \geq n \geq 10$ and $nD_{5/2}$ states with $5 \geq n \geq 16$ from Ref.~\cite{nistasd18}. We verified that the ionization potentials extracted from fitting $S$-series, shown in Fig.~\ref{fig:lines_and_fit}(d), and the $D$-series are in good agreement, see Tab. II. The ionization potential for Ca$^{+}$ extracted from the correlated fit using both data sets is 2~870~575.582(15)~GHz. The uncertainty stems from the fitting that includes errors in Tab. I. When we use only our results for high-lying $nS$ states together with fixed values for $\partial \mu/\partial E_{n,l,j}$ and $\mu^{1}_{l,j}$~\cite{djerad91a}, we obtain $I^{++}=$2~870~575.529(78)~GHz, with larger error, in full agreement with the previous result.

% and reducing largely the linear correlation in the residuals.

% This is the third way: We conjecture some systematic error in the data from~\cite{nistasd18} as a slight linear correlation in the fit residuals vanishes when these low-lying lines are shifted by $+6$~GHz, and $I^{++}=$2~870~575.536(20)~GHz is obtained.
\begin{table*}
\centering
\caption{Ionization energy and quantum defects calculated for the $nS_{1/2}$ and $nD_{5/2}$ states of $^{40}$Ca$^{+}$. The accepted value from Ref.~\cite{nistasd18} is given in the first row.}
\label{tab:fit_results}
{\renewcommand{\arraystretch}{1.4}
\begin{tabular}{m{4.8 cm } m{3.6 cm} m{2.1 cm} m{2.3 cm} m{1.9 cm} m{1.9 cm}} \hline \hline
Investigated states & $I^{++}$ [\si{\giga \hertz}] & $\mu^{0}_{l,j}$ & $R^* \partial \mu/\partial E_{n,l,j}$ & $\mu^{1}_{l,j}$ & Reference(s) \\
\hline
 & 2 870 568.8(9)  &  &  &  & \cite{nistasd18} \\
\hline
$nS_{1/2}$, $n$=5--10 &  2 870 577.3(7) & 1.803033(15) & $-0.1952(5)$ & 2.1127(26)  &   \\
$nS_{1/2}$, $n$=5--10,38--52,55,60,65   & 2 870 575.571(20) & 1.802995(5)  & $-0.1963(2)$ & 2.107(1) & this work \\
\hline
$nD_{5/2}$,$n$=5--16  & 2 870 577.4(8) & 0.62696(3)   & $-0.070(1)$  & 1.71(2)  &  \\
$nD_{5/2}$,$n$=5--16,37--50 & 2 870 575.596(23) &  0.626888(9)  & $-0.0723(5)$ & 1.671(9) & this work\\
\hline
Complete set $nS_{1/2}$, $nD_{5/2}$  & 2 870 575.582(15) &   &   &    & this work \\
\hline \hline
\end{tabular}
}
\end{table*}
%Previously measured low-lying transitions from~\cite{nistasd18} are included {\bf optionally ?? ich finde dass es aus der ersten spalte klar ist, welche n man nimmt, daher alles das weg}. The data set for $n$S$_{1/2}$ and $n$D$_{5/2}$ states is used either separately or collectively in the fitting procedure.

Our results for quantum defect coefficients of $nS_{1/2}$ and $nD_{5/2}$ series are given in Table~\ref{tab:fit_results}. These values are close to those which one can compute by fitting only to low-lying lines in~\cite{nistasd18}; however, the accuracy of $\mu_l$ has been improved by a factor of 3. We compared our values for quantum defects with those obtained from an accurate calculation of the Rydberg eigenenergies for Ca$^{+}$~\cite{pawlak20a} by calculating the energy differences from Eqn.~\ref{eq:ritz}, see supplementary material C. We calculate these energy differences below 750~MHz for the $nS_{{1/2}}$ series and down to 100~MHz for the $D_{5/2}$ series for the range of $n$ that we measured. This demonstrates that Rydberg spectroscopy is challenging theoretical calculations now at a much refined level.

%Another comparison with the values given in~\cite{djerad91a} shows \SIrange{0.8}{2}{\giga \hertz} energy differences between measured energies and those obtained from predicted quantum defects both $S$ and $D$ series.
%The ionization potential for Ca$^{+}$ from the fit is 2 870 575.986(16)~GHz, with the uncertainty calculated from a linear sum over one standard deviation from the fit and the systematic energy shifts in the measurements XXX. This result is 48 times more accuarate than the accepted value~\cite{nistasd18} and is $8\sigma$ above that, see Table~\ref{tab:fit_results}.
%Note that Rydberg series spectroscopy in which precise measurements of highly excited states lead to smaller contributions from energy-dependent terms of the quantum defect~(Eqn.~\ref{eq:qd}) is advantageous for determining the ionization energy. Note that ionization potentials extracted from independent fits to the $nS$ and $nD$ series are in good agreement with the result obtained from the correlated fit to the complete set of data.
%We also noticed that the ionization potential obtained only from a fit to low-lying energy terms for $nS$ and $nD$ states from~\cite{nistasd18} deviates $+9.5\sigma$ XXX better to give the value in GHz XXX from the accepted value.

In conclusion, we presented Rydberg spectroscopy of a single trapped Ca$^+$ to precisely determine parameters of the quantum defect theory. We improved the accuracy for quantum defects of $S$ and $D$-states and the ionization energy $I^{++}$, for which we suggest a new value. The corresponding modeling of specific shifts and broadening effects has been implemented and is successfully tested. Even high-lying Rydberg states are found to be stable in the dynamic Paul-potential. Such Rydberg states offer a novel platform with outstanding opportunities for tailoring strong dipolar state-dependent couplings, eventually  investigating correlated many-body systems. In future we will complement Rydberg spectroscopy by driving microwave transitions for exciting states with longer lifetimes, suited for fast quantum entangling operations in ion crystals~\cite{vogel19a, zhang20a}.

\begin{acknowledgments}
We thank TOPTICA for lending us the Wavemeter High Finesse WS8-2. This work was supported by the Deutsche Forschungsgemeinschaft (DFG) within the SPP 1929 Giant interactions in Rydberg Systems (GiRyd), the European Research Council under the European Union's Seventh Framework Programme (FP/2007-2013)  [ERC Grant Agreement No. 335266 (ESCQUMA)], within QuantERA by ERyQSenS, and by the EPSRC [Grant No. EP/R04340X/1]. A. M. acknowledges the funding from the European Union's Horizon 2020 research and innovation programme under the Marie Sk{\l}odowska-Curie grant agreement No.~796866 (Rydion).
\end{acknowledgments}

\bibliography{myreferences}

\clearpage

\begin{widetext}
\section*{Supplementary material: Precision measurement of the ionization energy of a single trapped $^{40}$Ca$^+$ ion by Rydberg series excitation}
\end{widetext}

\subsection{A. Ion loss probability}

\begin{figure}[!h]
  \includegraphics[width=0.48\textwidth]{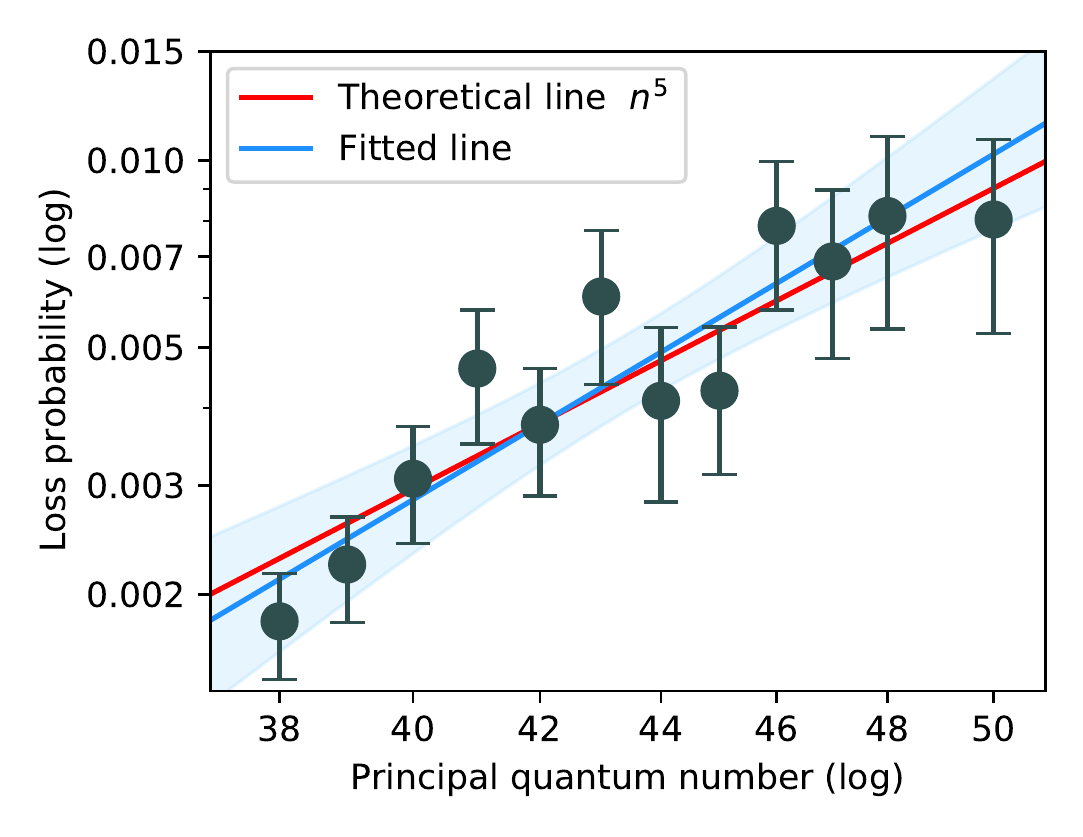}
  \caption{Loss probability data from fourth column of Table I plotted as a function of principle quantum number in log-log scale. We extract a power law scaling (blue) as $n^{5.7(0.9)}$, and show the 95$\%$ confidence band (light blue). The expected power law $n^{5}$ from ionization due to black-body radiation is plotted for comparison~[1] (red).}
\end{figure}

\subsection{B. Wavemeter accuracy cross-check}

\begin{figure*}
  \includegraphics[width=0.8\textwidth]{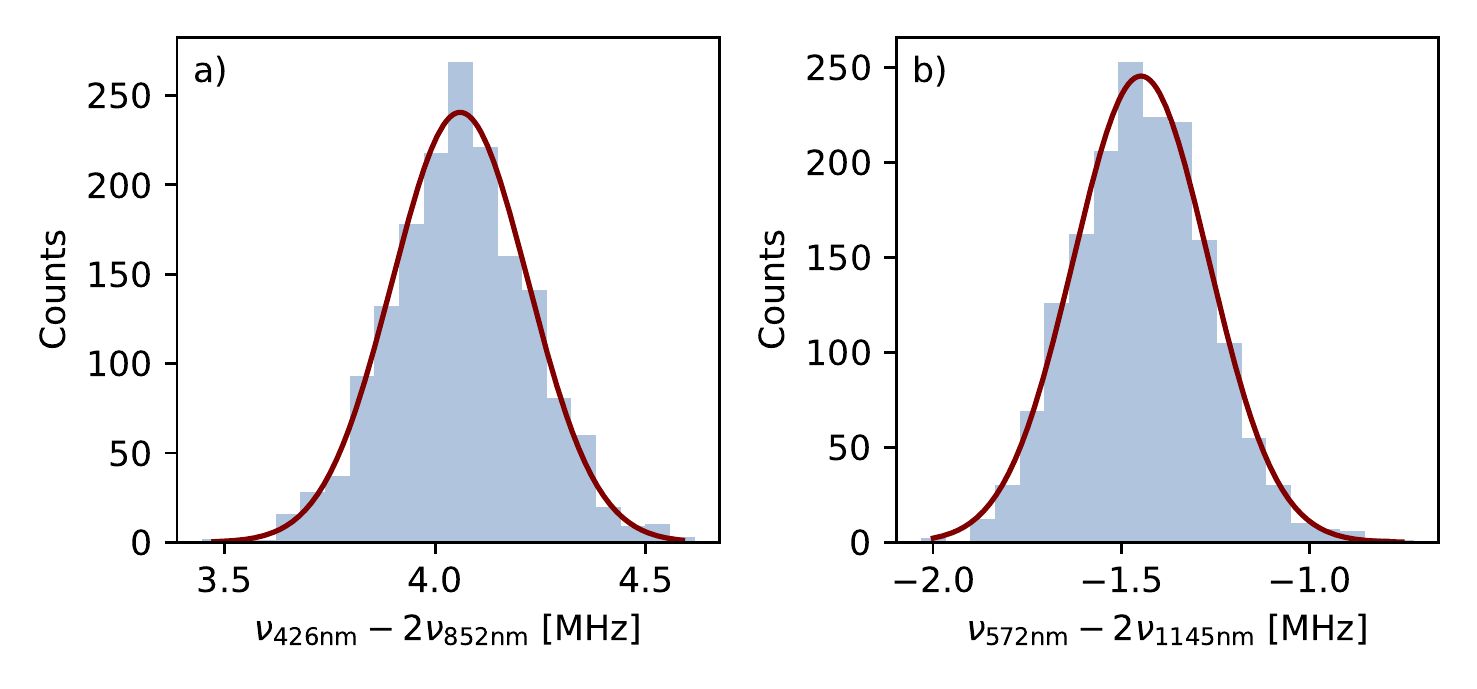}
  \caption{Histograms of $\nu_{\mathrm{SHG}} - 2 \cdot \nu_{\mathrm{fund}}$, where $\nu_{\mathrm{fund}}$ denotes frequency of laser fundamental frequency and $\nu_{\mathrm{SHG}}$ -- generated second harmonic frequency.
Both frequencies are measured by High Finesse WS8-2 wavemeter and each histogram contains 1680 measurements. a) Histogram for laser near 852~nm. b) Histogram for laser near 1145~nm. We fit the data to a Gaussian, revealing centers of +4.1~MHz and -1.4~MHz.
\label{histo_WM}}
\end{figure*}

The manufacturer states an absolute $3\sigma$ accuracy for the of WS8-2 of 10~MHz for wavelengths $\pm$200~nm around calibration wavelength and of 30~MHz outside this range~[2]. Wavelengths of lasers used to excite Rydberg states are measured at fundamental wavelengths 852~nm and 1145~nm. Thus, the $1\sigma$ accuracy results in 3.3~MHz and 10~MHz, respectively. To check this manufacturer-certified accuracy, we repeated 1680 measurements of both fundamental wavelengths (852~nm and 1145~nm) and compared them with that for the second harmonic (426~nm and 572~nm). For both comparisons the fundamental and frequency doubled wavelengths are in different accuracy ranges with respect to the calibration wavelength of  729~nm. The values were compared by taking a frequency difference of averaged values $\nu_\text{SHG} - 2 \times \nu_\text{fund}$, see Table~\ref{tab:freq_table} and fig.~\ref{histo_WM}. The shift of the histogram centers is  +4.1~MHz and -1.4~MHz confirming an accuracy within the stated specifications of the manufacturer. The reproducibility is extracted from the standard deviation of the histogram of measured differences results in 0.17~MHz and 0.18~MHz for 852~nm and 1145~nm respectively.

\begin{table*}
  \caption{Fundamental and second harmonic of laser systems near 213~nm and 287~nm measured with HighFinesse WS8-2 wavemeter. The histogram consists out of 1680 measurements. $\nu_\text{fund}$ denotes fundamental frequency, while $\nu_\text{SHG}$ denotes frequency of second harmonic.}
  \begin{tabular}{p{0.15\textwidth}p{0.15\textwidth}p{0.15\textwidth}p{0.20\textwidth}}
    \hline
    $\nu_\text{fund}$ [\si{MHz}] & $2 \times \nu_\text{fund}$ [\si{MHz}] & $\nu_\text{SHG}$ [\si{MHz}] & $\nu_\text{SHG} - 2 \times \nu_\text{fund}$ [\si{MHz}] \\
    \hline
    703 019 461.1 & 351 509 728.5 & 703 019 457.0 & 4.1 \\
    523 249 577.7 & 261 624 789.6 & 523 249 579.1 & -1.4\\
    \hline
  \end{tabular}
  \label{tab:freq_table}
\end{table*}

\subsection*{C. Comparison of calculated and measured quantum defects}

\begin{table}[!h]
\caption{The calculated values for the S and D quantum defect~[3] and the values revealed from the Rydberg spectroscopy data in Tab. II are used to determine the difference frequencies $\Delta \nu = \nu_{\text{exp.}} - \nu_{\text{theo.}}$. This difference accounts for about $1.9 \cdot 10^{-8}$ to $2.9 \cdot 10^{-7}$ of the Rydberg state energy for the  investigated range of principal quantum numbers $n$.}
  \begin{tabular}{p{0.07\textwidth}p{0.1\textwidth}p{0.1\textwidth}}
    \hline
    $n$ & $\Delta \nu_S$ [MHz] & $\Delta \nu_D$ [MHz] \\
    \hline
    37  & --     & 178.4  \\
    38  & -752.3 & 169.7  \\
    39  & -689.8 & 161.4  \\
    40  & -633.9 & 153.6  \\
    41  & -583.6 & 146.3  \\
    42  & -538.4 & 139.4  \\
    43  & -497.6 & 133.0  \\
    44  & -460.7 & 126.9  \\
    45  & -427.3 & 121.2  \\
    46  & -396.9 & 115.8  \\
    47  & -369.2 & 110.7  \\
    48  & -344.0 & 106.0  \\
    49  & -320.9 & 101.5  \\
    50  & -299.7 & 97.2   \\
    51  & -280.3 & --     \\
    52  & -262.5 & --     \\
    55  & -217.0 & --     \\
    60  & -161.2 & --     \\
    65  & -122.4 & --     \\
    \hline
  \end{tabular}
  \label{tab:QD_table}
\end{table}

[1] Quasiclassical calculations of blackbody-radiation-induced depopulation rates and effective lifetimes of {R}ydberg {$nS$}, {$nP$}, and {$nD$} alkali-metal atoms with {$n\ensuremath{\le}80$}, I. Beterov, I. Ryabtsev, D. Tretyakov, V. Entin, Phys. Rev. A {\bf 79} 052504 (2009)

[2] WS8-2 technical data, HighFinesse, {W{\"o}hrdstra{\ss}e 472072 T{\"u}bingen, Germany} (2020)

[3] Rydberg spectrum of a single trapped ${\mathrm{Ca}}^{+}$ ion: A Floquet analysis, M. Pawlak, H. Sadeghpour, Phys. Rev. A {\bf 101} 052510 (2020)

\end{document}